\documentclass[journal=nalefd,manuscript=letter]{achemso}

\usepackage{dcolumn}
\usepackage{bm}
\usepackage{xcolor}
\usepackage{xr-hyper}
\usepackage{amsmath,amsfonts,amssymb}
\usepackage{graphicx}
\usepackage{siunitx}

\usepackage{natmove}
\usepackage{hyperref}

\DeclareSIUnit[quantity-product = \,]
\degree{\text{deg}}

\makeatletter
\newcommand*{\addFileDependency}[1]{
  \typeout{(#1)}
  \@addtofilelist{#1}
  \IfFileExists{#1}{}{\typeout{No file #1.}}
}
\makeatother

\title{Nanoscale stray fields from micromagnets for optimal spin qubit architecture}

\author{S.~Lopes}
\affiliation{C12 Quantum Electronics, Paris, France}
\alsoaffiliation{Universit\'{e} de Lorraine, Institut Jean Lamour, UMR CNRS 7198, Nancy 54011, France}

\author{Q.~Schaeverbeke}
\affiliation{C12 Quantum Electronics, Paris, France}

\author{M.~M.~Desjardins}
\affiliation{C12 Quantum Electronics, Paris, France}

\author{D.~Lacour}
\affiliation{Universit\'{e} de Lorraine, Institut Jean Lamour, UMR CNRS 7198, Nancy 54011, France}

\author{M.~Hehn}
\affiliation{Universit\'{e} de Lorraine, Institut Jean Lamour, UMR CNRS 7198, Nancy 54011, France}

\author{F.~Montaigne}
\affiliation{Universit\'{e} de Lorraine, Institut Jean Lamour, UMR CNRS 7198, Nancy 54011, France}
\email{francois.montaigne@univ-lorraine.fr} 

\date{\today}

\begin{document}

%
\begin{abstract}

On-chip micromagnets generate local magnetic-field asymmetries, enabling electrical control of spin qubits via electric dipole spin resonance and their integration into circuit quantum electrodynamics (QED) architectures. Accurate prediction of spin-qubit performance requires modeling micromagnet stray fields beyond the saturated-magnet approximation, accounting for nonuniform magnetization. Here, we combine thin-film characterization of Co, Co/Ta multilayers, and CoFe films with nanoscale stray-field measurements using NV-center magnetometry in the unsaturated regime to establish a reliable micromagnetic simulation framework. We show that CoFe micromagnets generate antisymmetric fields in double quantum-dot geometries exceeding $\pm\SI{100}{\milli\tesla}$, owing to their high saturation magnetization and favorable magnetocrystalline anisotropy. For spin qubits coupled to microwave resonators, the predicted spin-photon coupling reaches $\left| g_s/g_c \right| \approx 0.5$, where $g_c$ denotes the charge-photon coupling strength of the underlying charge qubit, highlighting the potential for high-fidelity operations in circuit QED architectures.

\end{abstract}
\maketitle
%

%
%

%
Generating and controlling local magnetic fields and gradients at the nanometer length scale is experimentally challenging. Such control is sought across multiple research communities, including semiconductor spin qubits, topological quantum devices \cite{Zhou2019,Kornich2020,Turcotte2020,Jardine2021}, magnetic storage technologies \cite{INSIC2024Roadmap} and soft matter systems for nanoparticle manipulation \cite{Fratzl2013}. For spin-qubit applications, precise control of the magnetic-field distribution at these length scales is required to tailor the electronic spin energy spectrum. One approach to achieving such control is to use on-chip lithographically defined micromagnets to engineer the local magnetic distribution.

For electronic spins electrostatically confined in semiconductor quantum dots, engineered magnetic-field profiles are essential for coherent electrical spin control. A uniform magnetic field, effectively uniform on the scale of the individual quantum dot, is applied to lift the spin-state degeneracy, while a spatially varying magnetic-field gradient on the quantum-dot length scale is generated by a micromagnet. This gradient enables electrical control of spin transitions via electric dipole spin resonance (EDSR) \cite{Tokura2006, Pioro-Ladriere2008, Obata2010}. Using on-chip micromagnets, high-fidelity (99.9\%) single-qubit gates have been demonstrated on spin qubits confined in an electrostatically defined quantum dot with EDSR \cite{Yoneda2018}.

To enable fast, high-fidelity spin-qubit readout and long-range interactions between distant qubits, spin-qubit architectures have been extended from single quantum dots to double quantum dots coupled to microwave resonators \cite{Beaudoin2012, Mi2018, Samkharadze2018, Cubaynes2019}. In direct analogy to the single-dot case, magnetic field symmetric across the double quantum dot lifts the spin degeneracy, while an antisymmetric field between the two dots hybridizes the spin and charge degrees of freedom of the confined electron. This hybridisation endows the spin transition with an effective electric dipole moment, enabling efficient coupling to the electric field of a microwave resonator and realizing the so-called flip-flop spin qubit. Strong spin-photon coupling and coherent control with microwave cavities \cite{Mi2018, Samkharadze2018, Cubaynes2019}, as well as cavity-mediated interactions between distant spin qubits \cite{Borjans2020, Dijkema2025}, have been demonstrated. The coupling strength is set by the amplitude of the antisymmetric magnetic field, making it a central parameter for device performance \cite{Hu2012, Beaudoin2016, Benito2019}.

\begin{figure*}
    \centering
    \hspace{-0.45cm}\includegraphics[width=\linewidth]{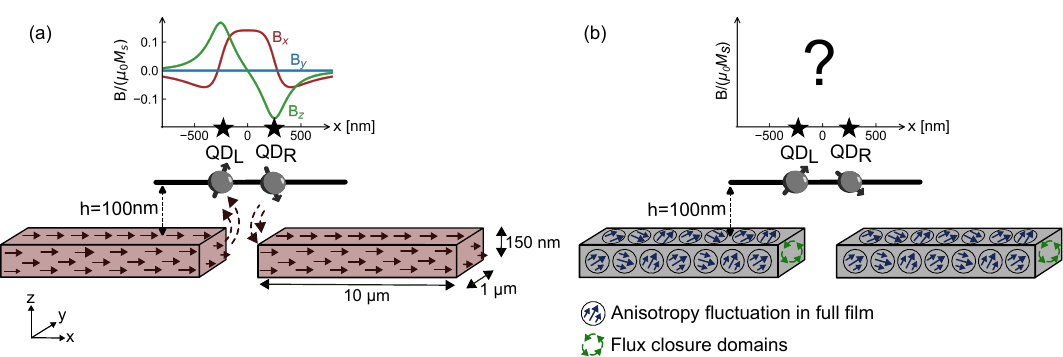}
    \caption{Optimal micromagnet geometry. The design consists of two elongated rectangular bars facing each other. (a) For uniform magnetization along the magnet long axis (x), the generated stray field is calculated; the geometry is optimized to maximize the antisymmetric component B$_\mathrm{z}$ with respect to the double quantum dot.  (b) Beyond the ideal uniform assumption, shape-induced dipolar interactions can generate flux-closure domains, and dispersion of magnetic anisotropy can further contribute to nonuniform magnetization. The elongated geometry exploits shape anisotropy to favor magnetization alignment along the long axis of the magnet.} 
    \label{fig:intro_magnet}
\end{figure*}

Micromagnet geometries producing magnetic-field gradients or local antisymmetric fields can be optimized under the assumption of uniform magnetization \cite{Legrand2023}. For a specified target magnetic configuration, the design is tailored to maximize the required gradient or field asymmetry. Here, we adopt a split pair of elongated rectangular micromagnets facing each other, optimized to maximize the antisymmetric magnetic-field component between two quantum-dot positions, based on the assumption of a uniform in-plane magnetization aligned with the magnet long axis (Fig.~\ref{fig:intro_magnet}). However, realistic on-chip micromagnets may deviate from uniform magnetization even under an applied field due to internal dipolar interactions and magnetic anisotropies \cite{Aldeghi2025NanoLett}. Accurately predicting the magnetization state, and thus the stray-field distribution, requires accounting for geometric effects that influence internal dipolar interactions. It also requires considering spatial variations of magnetic parameters within the material, particularly the dispersion of anisotropy directions. Reliable finite-difference micromagnetic modeling \cite{Vansteenkiste2014} therefore relies on precise knowledge of both material and structural properties. The relevant parameters include the magnetization at saturation, which sets the dipolar energy; the anisotropy constant, governing magnetocrystalline contributions; and structural features such as grain size and crystallographic-axis dispersion, which induce local anisotropy variations and reduce intergrain exchange coupling. The equilibrium magnetic configuration arises from the balance of these competing energy terms.

In this Letter, we combine experimental characterization and micromagnetic simulations to model the magnetization configuration and stray-field distribution of on-chip micromagnets made of cobalt-based materials in both saturated and unsaturated regimes. Thin-film measurements are first performed to extract the magnetic parameters used as input for the simulations. Local stray-field measurements are then conducted on patterned micromagnets using NV-center microscopy at low external magnetic fields, enabling direct probing of the magnetic field in the unsaturated regime. Owing to its high spatial resolution and Zeeman-shift-based readout, NV-center magnetometry enables quantitative nanoscale imaging of magnetic stray fields \cite{Rondin2013NatComm,Pelliccione2016NVscanning} and provides a direct benchmark of the micromagnetic simulation framework, which is then used to predict accurately spin-qubit properties.

%
%

\begin{figure*}
    \centering
    \hspace{-0.45cm}\includegraphics[width=\linewidth]{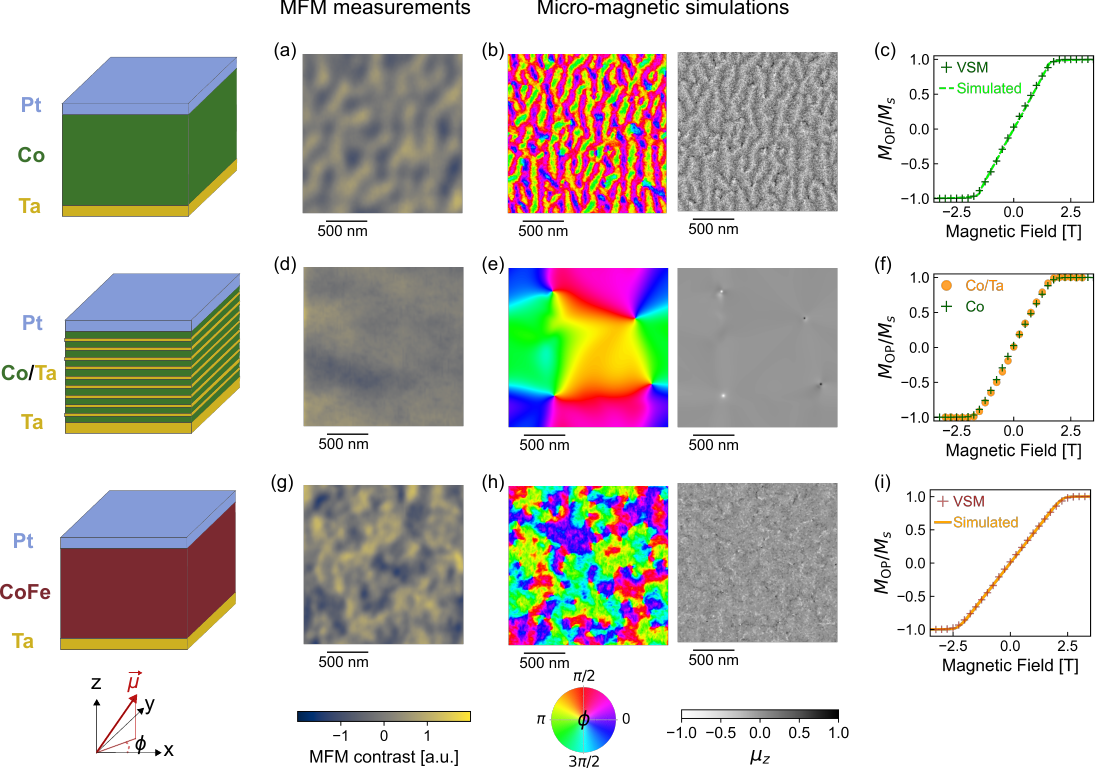}
    \caption{Thin-film measurements and micromagnetic simulations. (a) MFM scan above a Ta(\SI{5}{\nano \metre})/Co(\SI{150}{\nano \metre})/Pt(\SI{5}{\nano \metre}) film. (b) Simulated in-plane and out-of-plane magnetization at the film surface. (c) Out-of-plane vibrating sample magnetometry (VSM) of the Co film compared with micromagnetic simulations. (d)-(e) MFM measurements and simulated magnetization at the film surface for a Ta(\SI{5}{\nano \metre})/Co(\SI{15}{\nano \metre})/[Ta(\SI{1}{\nano \metre/Co(\SI{15}{\nano \metre})})]×9/Pt(\SI{5}{\nano \metre}) multilayer. (f) Out-of-plane VSM comparison between Co and Co/Ta films. (g)-(h) MFM measurements and simulated magnetization at the film surface for a Ta(\SI{5}{\nano \metre})/CoFe(\SI{150}{\nano \metre})/Pt(\SI{5}{\nano \metre}) film. (i) Out-of-plane VSM of the CoFe film compared with micromagnetic simulations.} 
    \label{fig:thin_film}
\end{figure*}

Three cobalt-based magnetic materials are investigated as candidates for the micromagnets. First, we consider a \SI{150}{\nano\metre}-thick cobalt (Co) film, one of the most commonly used materials for on-chip micromagnets \cite{Yoneda2018, Dijkema2025}. However, due to the magnetocrystalline anisotropy of hexagonal Co, complex three-dimensional stripe-like domain structures can form \cite{Hehn1996}, which may reduce and perturb the effective field generated by the magnet. This effect becomes particularly relevant for thicknesses typical of on-chip integration (\SIrange{100}{200}{\nano\metre}). To limit the emergence of such complex 3D configurations, we also investigate a cobalt-tantalum (Co/Ta) multilayer in which \SI{15}{\nano\metre}-thick Co layers are separated by \SI{1}{\nano\metre} amorphous tantalum spacers, limiting the development of structural and magnetic textures. In addition, we study a \SI{150}{\nano\metre}-thick cobalt-iron (CoFe) film. Its higher saturation magnetization enhances the stray field and dipolar interactions and, combined with the lower magnetocrystalline anisotropy of iron, favors predominantly in-plane magnetization compared to the Co film. However, the larger saturation magnetization also promotes the formation of in-plane, inhomogeneous magnetic domains.  The magnetic materials are deposited by sputtering onto Si wafers with a thermal oxide layer. An adhesion layer of \SI{5}{\nano \metre} of tantalum together with a capping layer of \SI{5}{\nano \metre} of platinum is added to the magnetic materials. Details on the material stack are provided in Fig. \ref{fig:thin_film}.

To gain initial insight into the remanent magnetization of the magnetic materials under study we perform magnetic force microscopy (MFM) measurements on the thin films after in-plane field saturation. The MFM contrast measured reflects magnetization inhomogeneities within the film volume or out-of-plane component of the magnetization at the surface.
As shown in Fig.~\ref{fig:thin_film}, the Co film exhibits a distinct stripe-like contrast, characteristic of an equilibrium configuration with magnetization pointing partially out of the film plane in alternating up and down orientations. In contrast, the MFM signal of the Co/Ta multilayer is strongly reduced. The CoFe film exhibits a higher contrast than the Co/Ta multilayer but without a well-defined domain pattern, indicating the presence of more in-plane inhomogeneous magnetic domains. \\
For the micromagnet design considered here, optimized for a saturated in-plane magnetization, the most suitable material is therefore one with predominantly in-plane magnetization and minimal domain inhomogeneity, i.e., exhibiting the lowest MFM contrast. Based on these measurements, the Co/Ta multilayer emerges as a promising candidate for on-chip micromagnet implementation. Nevertheless, all three materials are further characterized in the present study.

Magnetic properties of thin films are closely linked to their crystallographic structure through magnetocrystalline anisotropy (MCA), which defines preferred magnetization directions set by the crystal structure. X-ray diffraction (XRD) measurements, presented in the Supporting Information, reveal that the Co film adopts a hexagonal structure, known to exhibit strong uniaxial magnetocrystalline anisotropy (MCA) \cite{Cullity1972} favoring the formation of stripe domains consistent with the contrast observed in MFM.
This hexagonal Co structure is preserved in the Co/Ta multilayer, whereas the CoFe film exhibits a cubic structure. For all three films, the absence of several lattice planes in the diffraction patterns indicates textured films with a preferred crystallographic orientation rather than a randomly polycrystalline structure. \\
Micromagnetic simulations incorporate these structural features. Co and CoFe are modeled as granular materials. Magnetocrystalline anisotropy is included by assigning, within each grain, a uniaxial anisotropy axis for hexagonal structures or three orthogonal axes for cubic structures. To remain consistent with the XRD results, the anisotropy axes are not taken as randomly oriented but are constrained to lie along preferred crystallographic directions oriented following the crystallographic planes observed experimentally, with small grain-to-grain deviation. The Co/Ta film is modeled explicitly as a multilayer, where each Co layer forms a distinct magnetic region with reduced inter-layer exchange and no explicit grain structure. Additional details of the diffraction analysis and modeling are provided in the Supporting Information.

The magnetic properties of the thin films are further characterized using vibrating sample magnetometry (VSM), which measures the magnetic moment as a function of the applied field $B$. Normalization by the film volume yields the magnetization $M$. These measurements provide key quantitative parameters, in particular the saturation magnetization $M_s$ and the effective anisotropy field $B_K$.\\
Figure~\ref{fig:thin_film} shows the out-of-plane magnetization measured at room temperature with the magnetic field applied perpendicular to the film plane. The linear response observed for all three materials indicates that the easy axis of magnetization lies predominantly in the plane of the film. In-plane measurements, exhibiting hysteresis loops, are presented in the Supporting Information.\\
Assuming a magnetic thickness of \SI{150}{\nano\metre}, we extract the saturation magnetization values of $ M_s^\mathrm{Co} = \SI{1.45e6}{\ampere \per \metre}$,  $ M_s^\mathrm{Co/Ta} = \SI{9.3e5}{\ampere \per \metre}$ and $ M_s^\mathrm{CoFe} = \SI{1.88e6}{\ampere \per \metre}$. The values for Co and CoFe are consistent with literature reports \cite{Cullity1972}, whereas the saturation magnetization of the Co/Ta multilayer is reduced compared to pure Co, despite the same nominal magnetic thickness (\SI{150}{\nano\metre}). This reduction is likely due to magnetically dead layers at the Ta/Co interfaces, where interfacial disorder or intermixing can suppress the local magnetization.  In the micromagnetic simulations, the saturation magnetization is fixed to these measured values. 

Beyond the saturation magnetization, the effective anisotropy field $B_K$, defined as the hard-axis saturation field, can be extracted from the VSM data (Fig.~\ref{fig:thin_film}). In a macrospin approximation without magnetic anisotropy, one expects $B_K = \mu_0 M_s$, where $\mu_0$ is the vacuum permeability, whereas a perpendicular uniaxial anisotropy would reduce $B_K$ below this value. 
For the Co film, we obtain $B_K^\mathrm{Co} =\SI{1.85}{\tesla} \sim \mu_0 M_s^\mathrm{Co} = \SI{1.82}{T}$, indicating no clear evidence of magnetic anisotropy. In the simulations, a uniaxial anisotropy constant $K_{u,1}^\mathrm{Co} = \SI{500}{\kilo \joule \per \metre \cubed}$ at room temperature, taken from the literature \cite{Cullity1972}, reproduces the measured out-of-plane magnetization curve and, in particular, yields the same effective anisotropy field $B_K$ (Fig.~\ref{fig:thin_film}). The similar field dependence of the normalized out-of-plane magnetization curves for Co and Co/Ta (Fig.~\ref{fig:thin_film}) further indicates comparable intrinsic magnetic properties, consistent with a reduced effective magnetic volume in the Co/Ta multilayer due to magnetic dead layers.\\
In contrast, CoFe exhibits an enhanced anisotropy field, $B_K^\mathrm{CoFe} =\SI{2.6}{\tesla} > \mu_0 M_s^\mathrm{CoFe} = \SI{2.36}{T}$ consistent with cubic magnetocrystalline anisotropy. The simulations use a first-order cubic anisotropy constant $K_{c,1}^\mathrm{CoFe} = \SI{450}{\kilo \joule \per \metre \cubed}$ at room temperature to reproduce the VSM data (Fig.~\ref{fig:thin_film}). This exceeds the commonly reported value for Fe \cite{Cullity1972}, but is consistent with recent reports of enhanced cubic anisotropy in iron-based systems \cite{Aldeghi2025NanoLett}.\\ 
VSM measurements performed down to \SI{4}{\kelvin} further enable assessment of the temperature dependence of these magnetic parameters; low-temperature data are discussed in the Supporting Information.

Thin-film micromagnetic simulations yield equilibrium remanent configurations (see Fig.~\ref{fig:thin_film}) that can be directly compared with MFM measurements. The simulations reproduce stripe-like magnetization patterns for the Co film, consistent with the observed MFM contrast, and predict more inhomogeneous domain structures for CoFe than for the Co/Ta multilayer, in agreement with experimental observations. These results indicate that the magnetic properties of the thin films are accurately captured by the model.

The second step of the study focuses on the characterization and simulation of patterned micromagnets. At this scale, geometry and dipolar interactions strongly influence the magnetic configuration and may prevent full saturation even under an applied field. The accuracy of the simulation framework in capturing this behavior, including in the unsaturated regime, is assessed through quantitative measurements of the stray-field distribution generated by individual micromagnets.


%
%

NV-center magnetometry is particularly suited for this purpose as it combines high magnetic-field sensitivity with nanoscale spatial resolution at the level of a single micromagnet. A single nitrogen-vacancy (NV) center embedded near the apex of a diamond AFM tip acts as a point-like magnetic-field sensor \cite{Rondin2014}. The NV center is a spin-1 system whose spin-state-dependent photoluminescence enables optical detection of magnetic resonance (ODMR). When the NV center is subjected to an external magnetic field the degeneracy of the non-zero spin states is lifted, producing two resonance frequencies whose splitting is related to the local magnetic field. 

\begin{figure*}
    \centering
    \includegraphics[width=\linewidth]{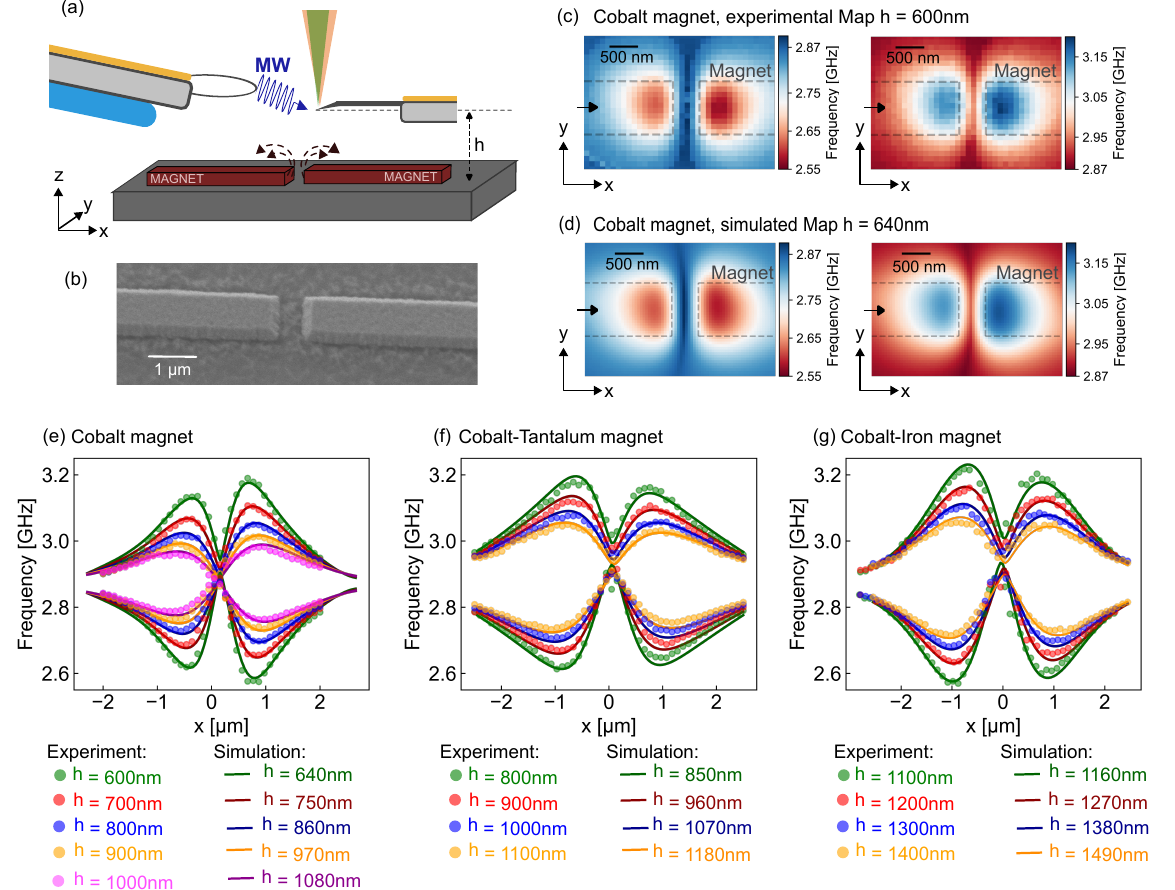}
    \caption{NV-center measurements and micromagnetic simulations.
(a) Schematic of the NV-center scanning probe microscope with a single NV center at the apex of an AFM cantilever, driven by a microwave field whose frequency is swept to induce spin transitions. 
(b) SEM image of a nanofabricated micromagnet consisting of two facing rectangular bars.
(c) Measured NV resonance frequencies at $h=\SI{600}{\nano\metre}$ above a Co micromagnet.
(d) Simulated NV resonance frequencies at $h=\SI{640}{\nano\metre}$ computed from micromagnetic stray-field simulations.
Dashed lines in (c) and (d) mark the micromagnet position.
(e) Line cuts of measured and simulated resonance frequencies along the arrow in (c) and (d).\\
(f),(g) Same as (e) for Co/Ta and CoFe micromagnets, respectively.
    }
    \label{fig:magnet_NV}
\end{figure*}

NV-center measurements performed above the micromagnet surface are shown in Fig.~\ref{fig:magnet_NV} for Co, Co/Ta, and CoFe micromagnets. The micromagnets are fabricated from thin-film magnetic materials using a titanium hard mask and a dry-etching process, ensuring CMOS compatibility. Two-dimensional scans are performed above the magnet surface. An ODMR spectrum is recorded at each position to extract the NV resonance frequencies.

The primary limitation of this technique is the maximum detectable magnetic field. Strong magnetic fields transverse to the NV axis - defined by the line connecting the nitrogen impurity to the adjacent vacancy - induce spin-state mixing under optical illumination, causing the spin-dependent photoluminescence contrast to vanish\cite{Rondin2014}. In practice, NV measurements become increasingly challenging for transverse fields $B_\perp \geq \SI{10}{\milli \tesla}$. To avoid this regime, the data presented here are acquired at relatively large stand-off distances, with the NV positioned at least \SI{600}{\nano \metre} above the sample surface.  While performing the measurements far from the sample surface inevitably blurs fine features of the stray-field distribution, the amplitude and spatial variation of the NV resonance frequencies still differ markedly between the three magnetic materials. This shows that, despite the loss of small-scale detail, the measurements retain sufficient information to allow a meaningful, quantitative comparison with the simulations.

In this study, rather than reconstructing the magnetic-field vector from the NV data \cite{Balasubramanian2008}, we use the micromagnetic simulations to compute the stray-field distribution of the patterned micromagnets. Room-temperature simulation parameters are used to match the measurement conditions, and the calculated stray fields are converted into NV resonance frequencies using the NV spin Hamiltonian (Supporting Information), enabling direct comparison with the experimental data, see Fig.~\ref{fig:magnet_NV}. Two-dimensional scans are performed at different stand-off heights and compared with simulations; representative line cuts extracted at selected heights are shown in Fig.~\ref{fig:magnet_NV}. To achieve optimal agreement, small offsets between the nominal experimental height and the effective simulation height are accounted for, primarily due to uncertainty in the NV position relative to the tip apex, variations in tip-sample engagement defining the $h=0$ plane, and piezoelectric scanner nonlinearities. The resulting height correction remains below 10\%.

%
%
%

The simulations, consistent with thin-film characterization, reproduce the room-temperature NV measurements, confirming the validity of the micromagnetic framework, including in the unsaturated regime. Applying the same VSM-based parameter extraction at low temperature (\SI{4}{\kelvin}) ensures its relevance under cryogenic conditions.

\begin{figure*}[t]
    \includegraphics[width=1\linewidth]{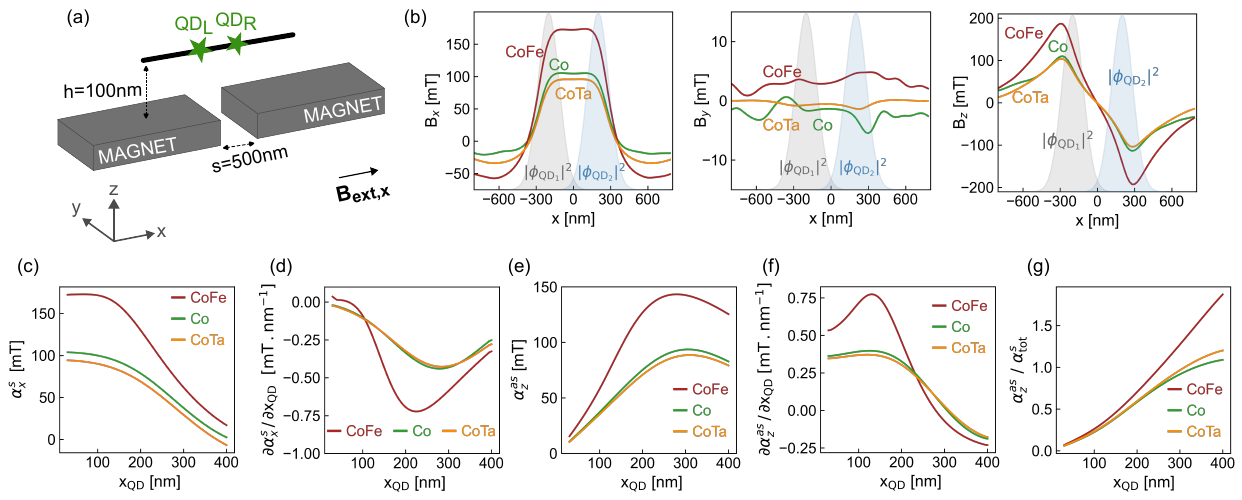}
    \caption{Predicted magnetic field along the double quantum-dot axis. (a) Schematic of the micromagnet and double quantum dot located at a height $h=\SI{100}{\nano\metre}$ above the magnet surface. The double quantum-dot axis is shown in black. (b) Magnetic-field components $B_\mathrm{x}$, $B_\mathrm{y}$, and $B_\mathrm{z}$ along the double quantum-dot axis. The dot positions are modeled by Gaussian probability densities centered at $\pm \mathrm{x}_\mathrm{QD}$ with full width at half maximum $\mathrm{x}_\mathrm{QD}$. (c) Effective symmetric field $\alpha_\mathrm{x}^{s}$, (d) its spatial derivative $\partial \alpha_\mathrm{x}^{s}/\partial \mathrm{x}_\mathrm{QD}$, (e) effective antisymmetric field $\alpha_\mathrm{z}^{as}$, (f) its spatial derivative $\partial \alpha_\mathrm{z}^{as}/\partial \mathrm{x}_\mathrm{QD}$, and (g) ratio $\alpha_\mathrm{z}^{as}/\alpha_\mathrm{tot}^{s}$, all shown as a function of $\mathrm{x}_\mathrm{QD}$.}
    \label{fig:mag_field_magnet}
\end{figure*}

The validated model is then used to simulate the magnetic-field distribution of the micromagnets and guide material selection. For integration within a double quantum dot architecture, the stray field is evaluated along the double quantum-dot axis using micromagnetic simulations with parameters extracted from thin-film measurements at \SI{4}{\kelvin}. The simulations are performed under an external field $B_\mathrm{ext,x} = \SI{50}{\milli \tesla}$, typical for spin-qubit operation \cite{Dijkema2025}. The results are presented in Fig.~\ref{fig:mag_field_magnet}.

For this geometry, the magnetization must be aligned with the long axis of the magnet (x direction in Fig.~\ref{fig:mag_field_magnet}) to maximize the anti-symmetric out-of-plane field $B_\mathrm{z}$. The micromagnet also generates a symmetric in-plane component $B_\mathrm{x}$. The external magnetic field applied along $\mathrm{x}$ further stabilizes uniform magnetization and adds a symmetric field contribution used to tune the spin-qubit energy spectrum.

For $B_\mathrm{ext,x} = \SI{50}{\milli \tesla}$, the micromagnets are not fully saturated and residual domains persist near the edges (see Supporting Information). This effect is most pronounced for Co, where stripe-like edge domains limit the stray-field amplitude. In contrast, the Co/Ta micromagnet produces a comparable antisymmetric field despite its lower effective saturation magnetization (see Fig.~\ref{fig:mag_field_magnet}). This behavior is attributed to reduced edge-domain formation in the Co/Ta multilayer, suggesting that further optimization, such as increasing the Co layer thickness while avoiding stripe formation at remanence, could enhance the effective saturation magnetization and the resulting generated antisymmetric field. \\
The largest stray-field amplitude is obtained with the CoFe micromagnet. Although its higher saturation magnetization would typically promote edge-domain formation, these domains are suppressed by the cubic magnetocrystalline anisotropy of CoFe (Supporting Information), resulting in an enhanced antisymmetric field. Based on this comparison, CoFe emerges as the most suitable magnetic material for the proposed magnet design.

To assess qubit properties, we consider the effective magnetic field experienced by each quantum dot, defined as
\begin{equation}
    \alpha_i^n =\int  \mathrm{d} \mathrm{x} ~B_i(\mathrm{x}) \left|\phi_{\mathrm{QD}_n}(\mathrm{x)}\right|^2
\end{equation}
 where $n\in \{ L,R\}$ labels the dots and $i\in \{ \mathrm{x,y,z}\}$ the field components. The electron probability density $\left|\phi_{\mathrm{QD}_n}\right|^2$  is modeled as a Gaussian centered at $\pm~\mathrm{x}_\mathrm{QD}$ with a full width at half maximum equal to $\mathrm{x}_\mathrm{QD}$, ensuring finite overlap between the electron densities of the two dots (see Fig. \ref{fig:mag_field_magnet}). 

 \noindent The qubit properties are determined by the symmetric field $\alpha_\mathrm{x}^s = (\alpha_\mathrm{x}^L +\alpha_\mathrm{x}^R)/2$ and the antisymmetric field $\alpha_\mathrm{z}^{as} = (\alpha_\mathrm{z}^L -\alpha_\mathrm{z}^R)/2 $.
The values of these fields depend on the dot positions and are presented in Fig. \ref{fig:mag_field_magnet}.

The key figures of merit of the qubit are its addressability and coherence, which are governed by the magnetic-field distribution at the quantum-dot positions. Balancing the benefits of strong spin-charge hybridization with decoherence contributions, a desirable regime is achieved when $\alpha_\mathrm{z}^{as} \sim \alpha^s_\mathrm{tot}$ where $ \alpha_{\mathrm{tot}}^s =\alpha_\mathrm{x}^s + B_{\mathrm{ext},\mathrm{x}}$ (see Supporting Information). In addition, a tradeoff between tunability and dephasing must be considered. Spatial variations of the magnetic field at the quantum-dot scale convert charge-noise-induced position fluctuations into qubit dephasing. Minimizing the spatial derivatives of the magnetic field therefore enhances resilience to charge noise (see Fig.~\ref{fig:mag_field_magnet}). Conversely, such spatial variations are required for scalable architectures, as they enable all-electrical tuning of the qubit frequency via electrostatic displacement of the dot, allowing local qubit addressability.
 
\begin{figure*}[t]
    \includegraphics[width=1\linewidth]{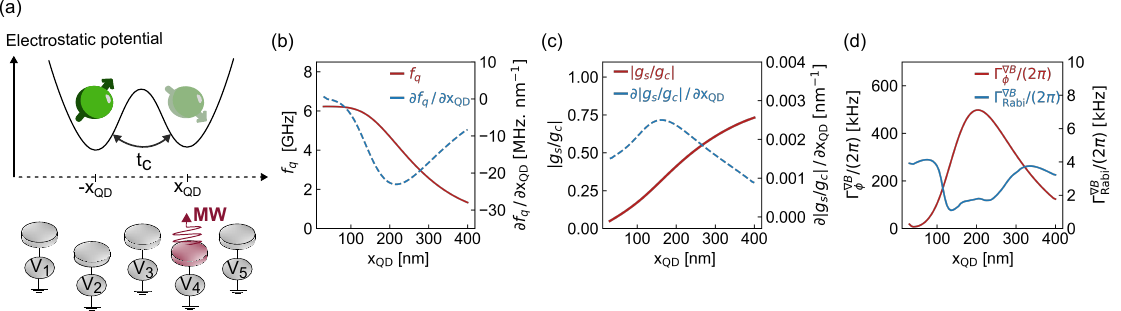}
    \caption{Prediction of qubit properties from the CoFe micromagnet stray-field distribution. (a) Schematic of a gate-defined double quantum dot (DQD) hosting a single electron; t$_\mathrm{c}$ denotes the interdot tunnel coupling. For the simulations shown here, $t_c = \mu_B g \alpha^s_\mathrm{tot}$ where $\mu_B$ is the Bohr magneton and $g\approx2$ is the electron Landé $g$-factor. The spin properties are governed by the magnetic-field distribution generated by a CoFe micromagnet (not represented). Spin transitions are driven electrically by applying a microwave (MW) tone to a nearby gate. (b) Simulated qubit resonance frequency and its spatial derivative as a function of the dot position x$_\mathrm{QD}$. (c) Spin–photon coupling strength $\left| g_s \right|$ and its spatial derivative versus x$_\mathrm{QD}$, where $\left| g_s \right|$ depends on the charge-photon coupling strength $\left| g_c \right|$, defined for a purely charge qubit. (d) Predicted qubit dephasing rate $\Gamma_\phi^{\nabla B}$ arising from charge-noise-induced position fluctuations, assuming quasi-static charge noise. The corresponding Rabi decoherence rate, $\Gamma_\mathrm{Rabi}^{\nabla B}$, under MW driving is also shown. }
    \label{fig:mag_geo_comp}
\end{figure*}

To assess the figure of merit of the qubit, we compute the energy spectrum of a single electron confined in a double quantum dot and subjected to the effective magnetic fields $\alpha_i^n $ for $i \in \{\mathrm{x,y,z}\} $ and $n \in \{L,R\}$.  We focus on the CoFe micromagnet due to its higher magnetic field amplitudes and improved magnetic configuration homogeneity. From this spectrum, we extract the qubit frequency $f_q$, its spatial derivative, the spin-photon coupling strength, characterized by the ratio $\left|g_s/g_c\right| $ and its spatial derivative, as well as the resulting qubit dephasing rate (Fig.~\ref{fig:mag_geo_comp}). Details of the derivations are provided in the Supporting Information.

For the present geometry, the antisymmetric field is maximized by positioning the quantum dots near the magnet edges. This enhances spin–charge hybridization, enabling electrical spin control and coupling to microwave resonators for dispersive readout and long-range interactions. At these positions, the qubit frequency exhibits a finite spatial dependence, which enables all-electrical tunability required for scalable architectures. However, the same spatial variation converts charge-noise-induced position fluctuations into qubit dephasing. The optimal dot position therefore results from a balance between electrical controllability and coherence and depends on the charge-noise environment of the platform. Figure~\ref{fig:mag_geo_comp} shows the predicted dephasing and Rabi decoherence rates assuming charge-noise levels of Si/SiGe spin-qubit devices reported in the literature \cite{Kawakami2014LongLivedSpinQubit}. In this analysis, we consider only the effect of charge noise that modulates the dot position within the spatially varying magnetic field, focusing on the decoherence arising from the micromagnet field profile. Other charge-noise effects acting directly on the charge degree of freedom are not included, as they are not specific to the micromagnet. Further details of the derivation are provided in the Supporting Information. A similar operating regime can be achieved by adapting the micromagnet geometry, rather than varying the dot position $\mathrm{x}_\mathrm{QD}$ relative to the magnet edges (see Supporting Information).


%

In conclusion, thin-film magnetic characterization is combined with micromagnetic simulations to quantitatively model the stray fields generated by on-chip micromagnets made of Co, Co/Ta multilayers, and CoFe. Local NV-center measurements on patterned micromagnets are in agreement with simulations, establishing a reliable framework applicable in both saturated and unsaturated regimes. The sputtered magnetic films exhibit a textured structure with well-defined preferred crystallographic orientations, reducing polycrystallinity-induced stray-field variations associated with random dispersion of magnetocrystalline anisotropy \cite{Aldeghi2025NanoLett}. Among the materials investigated, CoFe emerges as the most suitable for double quantum-dot spin qubits, enabling large antisymmetric fields due to its high saturation magnetization and favorable anisotropy.\\
Using this framework, we simulate the magnetic-field distribution generated by a CoFe micromagnet and predict the resulting qubit properties. The magnet geometry enhances spin-charge hybridization, enabling electrical spin control and resonator-mediated manipulation, while the controlled spatial magnetic field variation opens a path to all-electrical frequency tunability essential for scalable architectures. These benefits, however, must be balanced against coherence, as strong hybridization and finite field spatial derivatives increase sensitivity to charge noise. Our analysis shows that a favorable regime with $g_s \gg \Gamma^{\nabla B}$, enabling high-fidelity coherent electrical manipulation via a microwave resonator, can be achieved using high-kinetic-inductance resonators that enhance the charge–photon coupling to several hundred $\si{\mega \hertz}$ \cite{Samkharadze2018,Yu2023}.\\
Beyond double quantum-dot architectures, the characterization and modeling approach developed here is broadly applicable. In single quantum-dot platforms, accurate knowledge of the micromagnet stray-field distribution is essential for electrical spin control, and in larger quantum-dot arrays, engineered field profiles enable individual qubit addressability \cite{Philips2022Nature} and single-qubit gate operations via hopping gates \cite{Unseld2025Baseband}.


\begin{acknowledgement}
This work was supported partly by France 2030 government investment plan managed by the French National Research Agency under grant references Equipex e-DIAMANT ANR-21-ESRE-0031, PEPR SPIN-ANR-22-EXSP0007 and PEPR MIRACLEQ ANR-23-PETQ-0003.
\end{acknowledgement}

%


\nocite{*}
\bibliography{biblio}
\end{document}